\documentclass[aps,prd,preprint,superscriptaddress,tightenlines,nofootinbib]{revtex4}

\usepackage{rotating}
\usepackage{graphicx}
\usepackage{epsfig}
\usepackage{graphicx}
\usepackage{bm}

\newcommand{\dprhoe}{D^+ \to \rho^0 e^+ \nu_e}

\newcommand{\dpke}{D^+ \to \bar{K}^0e^+\nu_e}

\newcommand{\dpkste}{D^+ \to \bar{K}^{*0}e^+\nu_e}

\newcommand{\dppie}{ D^+ \to \pi^0e^+\nu_e}

\newcommand{\dpome}{ D^+ \to \omega e^+\nu_e}

\newcommand{\vcs}{V_{cs}}
\newcommand{\vcd}{V_{cd}}

\begin{document}


\preprint{CLNS 05-1915}       
\preprint{CLEO 05-07}         

\title{ Absolute Branching Fraction Measurements of  Exclusive
        $D^+$ Semileptonic Decays}

\author{G.~S.~Huang}
\author{D.~H.~Miller}
\author{V.~Pavlunin}
\author{B.~Sanghi}
\author{I.~P.~J.~Shipsey}
\affiliation{Purdue University, West Lafayette, Indiana 47907}
\author{G.~S.~Adams}
\author{M.~Chasse}
\author{M.~Cravey}
\author{J.~P.~Cummings}
\author{I.~Danko}
\author{J.~Napolitano}
\affiliation{Rensselaer Polytechnic Institute, Troy, New York 12180}
% ...
\collaboration{CLEO Collaboration} 
\noaffiliation


\date{\today}

\begin{abstract}

Using data collected at the $\psi(3770)$ resonance with the CLEO-c
detector at the Cornell $e^+ e^-$ storage ring, we present 
improved measurements of the absolute branching fractions of $D^+$ 
decays to $\bar{K}^0 e^+ \nu_e$, $\pi^0 e^+ \nu_e$, $\bar{K}^{*0} e^+ \nu_e$,
and $\rho^0 e^+ \nu_e$, and the first 
observation and absolute branching fraction measurement of $\dpome$.
We also report the most precise tests to date of isospin invariance in 
semileptonic $D^0$ and $D^+$ decays.

\end{abstract}


\maketitle

The quark mixing parameters are fundamental constants of 
the Standard Model~(SM) of particle physics. They determine 
the nine weak-current quark coupling elements of the 
Cabibbo-Kobayashi-Maskawa~(CKM) matrix~\cite{ckm}. 
The extraction of the quark couplings is difficult because 
quarks are bound inside hadrons by the strong interaction. 
Semileptonic decays are the preferred way to determine the
CKM matrix elements as the strong interaction binding effects 
are confined to the hadronic current. They are parameterized 
by form factors that are calculable,  
for example, by lattice quantum chromodynamics~(LQCD) 
and QCD sum rules. Nevertheless, 
form factor uncertainties dominate the precision with which 
the CKM matrix elements 
can be determined~\cite{VubVcb}.  
In charm quark decays, however, couplings 
$\vcs$ and $\vcd$ are tightly constrained by the unitarity 
of the CKM matrix. Therefore, measurements of charm 
semileptonic decay rates and form factors rigorously 
test theoretical predictions. 

We report herein measurements with the first 
CLEO-c data~\cite{cleoc} of the absolute branching fractions of 
$D^+$ decays to  $\bar{K}^0 e^+ \nu_e$, $\pi^0 e^+ \nu_e$, 
$\bar{K}^{*0} e^+ \nu_e$, and $\rho^0 e^+ \nu_e$,
and the first observation and absolute branching fraction 
measurement of $\dpome$. 
(Throughout this Letter charge-conjugate modes are implied.)
We combine these results with the measurements of $D^0$ semileptonic 
branching fractions reported in Ref.~\cite{cleoc-neutral_semilep}, which 
use the same data and analysis technique, and test isospin invariance 
of the hadronic current in semileptonic decays.

The data were collected by the CLEO-c detector at the 
$\psi (3770)$ resonance, about 40~MeV above 
the $D \bar{D}$ pair production threshold.
A description of the CLEO-c detector is provided in Ref.~\cite{cleoc-neutral_semilep} 
and references therein. The data sample consists of an integrated
luminosity of 55.8~pb$^{-1}$ and includes about 0.16~million $D^+ D^-$
events.

The technique for these measurements was first applied by the Mark III
collaboration~\cite{MkIII} at SPEAR. Candidate events
are selected by reconstructing a $D^-$, called a tag, in the
following six hadronic final states: 
$K^0_S \pi^- $, $K^+ \pi^- \pi^-$, $K^0_S \pi^- \pi^0 $, 
$K^+ \pi^- \pi^- \pi^0$, $K^0_S \pi^- \pi^- \pi^+$, and $K^- K^+ \pi^- $. 
The absolute branching fractions of $D^+$ semileptonic decays are then 
measured by their reconstruction in the system recoiling from the tag.
Tagging a $D^-$ meson in a $\psi(3770)$ decay provides a $D^+$ 
with known four-momentum, allowing a semileptonic decay to be 
reconstructed with no kinematic ambiguity, even though the neutrino 
is undetected.

Tagged events are selected based on two variables: $\Delta E
\equiv E_{D} - E_{\rm beam}$,  the difference between the energy 
of the $D^{-}$ tag candidate ($E_{D}$) and
the beam energy ($E_{\rm beam}$), and the beam-constrained mass 
$M_{\rm bc} \equiv \sqrt{ E_{\rm beam}^2/c^4 - |\vec{p}_{D}|^2/c^2}$, 
where $\vec{p}_{D}$ is the measured momentum of the $D^{-}$ candidate.
Selection criteria for tracks,  $\pi^{0}$ and $K^0_S$  candidates
for tags are described in Ref.~\cite{cleoc-Dtagging}. If multiple
candidates are present in the same tag mode, one candidate 
per tag charge is chosen using $\Delta E$. The yields
of the six tag modes are obtained from fits to the $M_{\rm bc}$
distributions. The data sample comprises approximately 32,000 
charged tags~(Table~\ref{table1}).

\begin{table}
\begin{center}
\begin{tabular}{lc}
\hline
\hline
Tag Mode &  Yield  \\ 
\hline
$D^- \rightarrow  K^0_S \pi^- $         &   $2243 \pm 51 $ \\  
$D^- \rightarrow  K^+ \pi^- \pi^-$   &   $15174 \pm 128 $ \\   
$D^- \rightarrow  K^0_S \pi^- \pi^0 $  &   $5188 \pm 100 $  \\ 
$D^- \rightarrow  K^+ \pi^- \pi^- \pi^0$    &   $4734 \pm 91$ \\ 
$D^- \rightarrow  K^0_S \pi^-  \pi^- \pi^+$   &   $3281 \pm 94$ \\ 
$D^- \rightarrow  K^- K^+ \pi^- $           &   $1302 \pm 44$ \\   
\hline
All Tags & $31922 \pm 219$ \\
\hline
\hline
\end{tabular}
\end{center}
\caption{Tag yields of the six  $D^{-}$ hadronic modes with 
         statistical uncertainties. }
\label{table1}
\end{table}

After a tag is identified, we search for a positron
and a set of hadrons recoiling against the tag.
(Muons are not used as $D$ semileptonic decays at the $\psi(3770)$
produce low momentum leptons for which the CLEO-c muon
identification is not efficient.)
Positron candidates, selected with criteria described 
in Ref.~\cite{cleoc-neutral_semilep}, 
are required to have momentum of at least
200~MeV/$c$ and to satisfy $| \cos{\theta} |$ $<$ 0.90, where
$\theta$ is the angle between the positron direction and
the beam axis. The efficiency for positron identification rises from about
$50\%$ at 200~MeV/$c$ to 95\% just above 300~MeV/$c$ and is roughly constant
thereafter.  The rates for misidentifying charged pions and kaons
as positrons averaged over the momentum range are approximately 0.1\%.
Bremsstrahlung photons are recovered by the procedure 
described in Ref.~\cite{cleoc-neutral_semilep}.

Hadronic tracks must have momenta above 50~MeV/$c$ and
$| \cos{\theta} | < 0.93$. Identification of hadrons
is based on measurements of specific ionization~($dE/dx$)
in the main drift chamber and information from the Ring
Imaging Cherenkov Detector~(RICH). Pion and kaon candidates are
required to have  $dE/dx$ measurements within three standard
deviations (3.0$\sigma$) of the expected value. 
For tracks with momenta greater than 700~MeV/$c$, RICH information,
if available, is combined with $dE/dx$.  The efficiencies ($95$\% or
higher) and misidentification rates (a few per cent) 
are determined with charged pion and kaon samples from 
hadronic $D$ decays.

We form $\pi^0$ candidates from pairs of photons, each having an
energy of at least 30 MeV, and require that the invariant mass of
the pair be within 3.0$\sigma$ ($\sigma \sim 6$~${\rm MeV}/c^2$)  
of the known $\pi^0$ mass. A mass
constraint is imposed when $\pi^0$ candidates are used in further
reconstruction. The $K^0_S$ candidates are formed from pairs of
oppositely-charged and vertex-constrained tracks having an
invariant mass within 12 MeV/$c^2$ $( \sim 4.5 \sigma )$ of the 
known $K^0_S$ mass. We form a $\bar{K}^{*0}$~($\rho^0$) candidate 
from $K^-$ and $\pi^+$~($\pi^-$ and $\pi^+$) candidates and require 
an invariant mass within 100~MeV/$c^2$~(150~MeV/$c^2$)
of its mean value. The reconstruction 
of $\omega \rightarrow \pi^+ \pi^- \pi^0$ candidates is achieved by
combining three pions, requiring an invariant mass within
20~MeV/$c^2$ of the known mass, and demanding that the charged
pions do not satisfy interpretation as a ${K}^0_S$.

The tag and the semileptonic decay are then combined,
if the event includes no tracks other than those of the tag
and the semileptonic candidate.
Semileptonic decays are identified using the variable 
$U \equiv E_{\rm miss} - c|\vec{p}_{\rm miss}|$, where $E_{\rm miss}$
and $\vec{p}_{\rm miss}$ are the missing energy and momentum of
the $D$ meson decaying semileptonically. If the decay products of 
the semileptonic decay have been correctly identified, 
$U$ is expected to be zero,  since only a neutrino is undetected. 
The resolution in $U$ is improved using constraints described
in Ref.~\cite{cleoc-neutral_semilep}. 
Due to the finite resolution of the detector, the distribution in $U$ 
is approximately Gaussian, centered at $U=0$ with 
$\sigma \sim 10~{\rm MeV}$. (The width varies by mode and is
larger for modes with neutral pions.)  
To remove multiple candidates in each semileptonic mode, 
one combination is chosen per tag mode, based on the 
proximity of the invariant masses of the $K_S^0$,  $\bar{K}^{*0}$, 
$\rho^0$, $\pi^0$, or $\omega$ candidates to their expected masses.

\begin{figure}
\epsfig{file=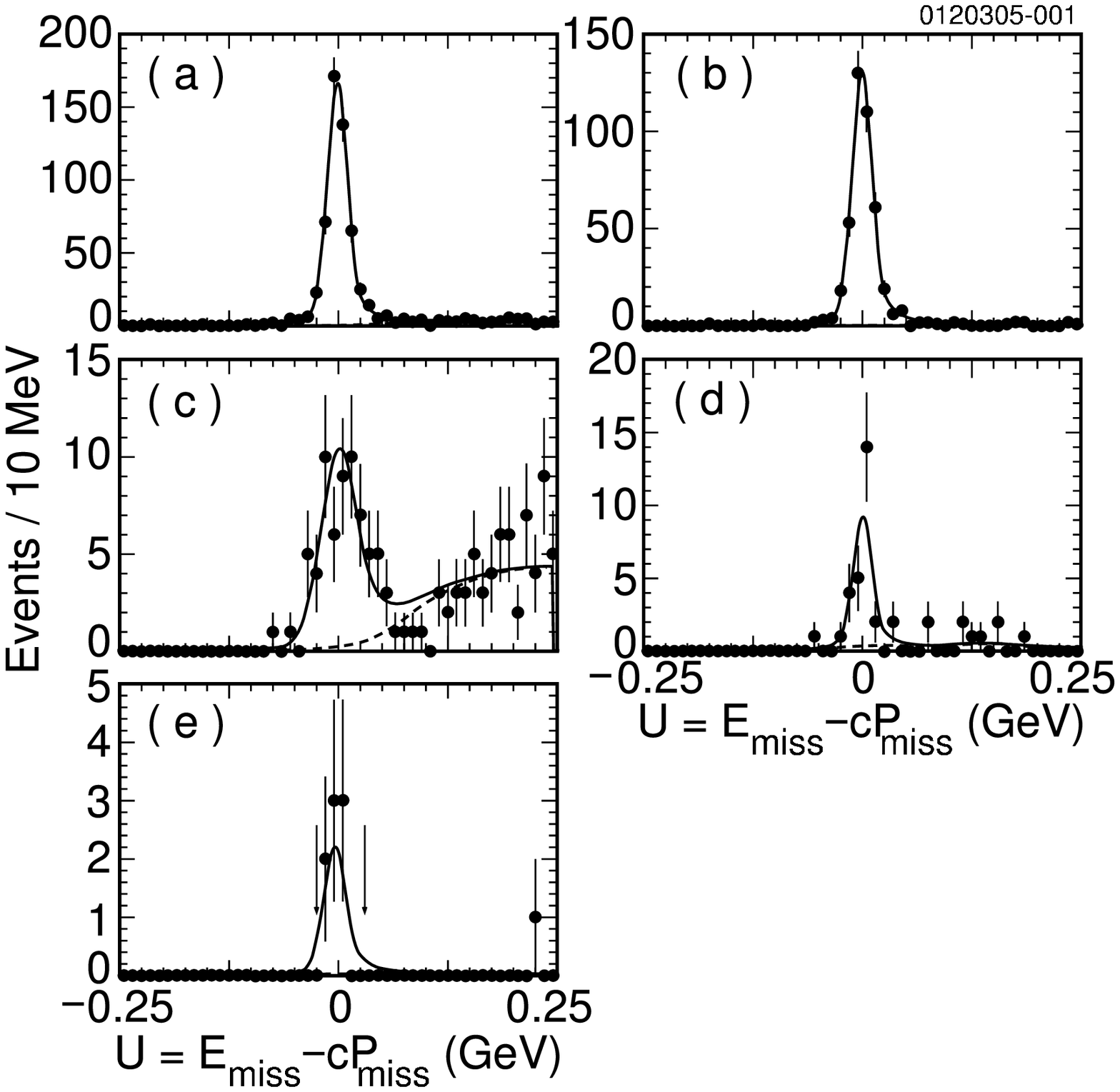,width=3.6in} \caption{Fits~(solid
line) to the $U$ distributions in data~(dots with error bars) for
the five $D^+$ semileptonic modes: (a)~$\dpke$, (b)~$\dpkste$,
(c)~$\dppie$, (d)~$\dprhoe$ and (e)~$\dpome$. The background
contribution is the dashed line only visible in (c) and (d). 
The arrows show the signal region for $\dpome$.
}
\label{figure1}
\end{figure}

The yield for each semileptonic mode is determined from a fit to
its $U$ distribution, as shown in Fig.~\ref{figure1} with all tag 
modes combined. In each case the signal is represented by a Gaussian 
and a Crystal Ball function~\cite{cb} to account for initial and
final state radiation~(FSR). The parameters describing the tails of the
signal function are fixed with 
a GEANT-based Monte Carlo~(MC) simulation~\cite{GEANT}. The background 
functions are determined by a MC simulation that incorporates all available
data on $D$ meson decays.
The backgrounds are small and arise mostly from misreconstructed
semileptonic decays with correctly reconstructed tags~\cite{rhoenu}.
The background shape parameters are fixed, while the background
normalizations are allowed to float in all fits to the data.

The mode $D^+ \rightarrow \omega e^+ \nu_e $ has never previously
been observed. There are 8 events consistent with $D^+
\rightarrow \omega e^+ \nu_e $ in Fig.~\ref{figure1}~(e).
The background in the signal region~($[-25;+30]$~MeV in~$U$)
is estimated to be $0.4\pm 0.2$ events.
The probability for the background of 0.6~events to fluctuate to
8 or more events is $2.4 \times 10^{-7}$, which
corresponds to significance exceeding $5.0 \sigma$. Therefore,
this is the first observation of $D^+ \rightarrow \omega e^+ \nu_e $.

The absolute branching fractions in Table~\ref{table2} are determined 
using $ {\cal B}={N_{\rm signal} / \epsilon N_{\rm tag}}$,
where $N_{\rm signal}$ is the number of fully reconstructed $D^{+} D^{-}$
events obtained by fitting the $U$ distribution,
$N_{\rm tag}$ is the number of events with a reconstructed tag,
and $\epsilon$ is the effective efficiency for detecting the semileptonic
decay in an event with an identified tag. A MC simulation
where the relative population of tag yields across tag modes
approximates that in the data is used to determine the efficiency.

\begin{table*}
\begin{center}
\begin{tabular}{ l c c c c}
\hline \hline
 Decay Mode &  \hspace{5mm} $\epsilon$ (\%) \hspace{5mm}  & \hspace{5mm} Yield \hspace{5mm}  & \hspace{5mm} $\mathcal{B}$ (\%) \hspace{5mm}  &   \hspace{2mm} $\mathcal{B}$ (\%) (PDG) \hspace{2mm} \\
\hline
$D^+ \rightarrow  \bar{K}^0 e^+ \nu_e  $ & $57.1 \pm 0.4$ & $545 \pm 24 $
                                      & $ 8.71 \pm 0.38 \pm 0.37 $  &  $6.7  \pm 0.9$        \\
$D^+ \rightarrow  \pi^0 e^+ \nu_e  $ & $45.2 \pm 1.0$ & $63.0 \pm 8.5$
                                      & $ 0.44 \pm  0.06 \pm 0.03 $  &  $0.31 \pm 0.15$   \\
$D^+ \rightarrow  \bar{K}^{*0} e^+ \nu_e  $ & $34.8 \pm 0.3$
                                     & $422 \pm 21$   & $5.56 \pm 0.27 \pm 0.23 $  &  $5.5  \pm 0.7$ \\
$D^+ \rightarrow  \rho^0 e^+ \nu_e  $ & $40.0 \pm 1.1$
                                     & $27.4 \pm 5.7$   & $ 0.21 \pm 0.04 \pm 0.01 $ & $0.25 \pm 0.10$ \\
$D^+ \rightarrow  \omega e^+ \nu_e  $ & $16.4 \pm 0.6$ & $ 7.6^{+3.3}_{-2.7}$
                                      &  $ 0.16^{+0.07}_{-0.06} \pm 0.01 $ &  ---    \\
\hline \hline
\end{tabular}
\end{center}
\caption{Signal efficiencies, yields, and branching fractions
         in this work and a comparison to the PDG~\cite{PDG}.
         The first uncertainty is statistical and the second 
         systematic in the fourth column, and 
         statistical or total in the other columns.
         The efficiencies do not include subsidiary branching
         fractions. The $D^+ \rightarrow  \bar{K}^{*0} e^+ \nu_e$ yield 
         is reduced by 2.4\% in the calculation of the branching 
         fraction~(see the text for details). }
\label{table2}
\end{table*}

We have considered the following sources of systematic uncertainty
and give our estimates of their magnitudes in parentheses. The
uncertainties associated with the efficiency for finding a track
(0.7\%), $\pi^0$ (2.0\% for $D^+ \rightarrow \omega e^+ \nu_e$ and 4.3\%
for $D^+ \rightarrow \pi^0 e^+ \nu_e$)
and $K_S^0$~(3.0\%) are estimated
using missing mass techniques with the data~\cite{cleoc-Dtagging}.
Details on the uncertainties associated with positron identification
efficiency (1.0\%) are provided in Ref.~\cite{cleoc-neutral_semilep}.
Uncertainties in the charged pion and kaon
identification efficiencies~(0.3\% per pion and 1.3\% per kaon) 
are estimated using hadronic $D$ meson decays. 
The uncertainty in the
number of tags~(1.1\%) is estimated by using alternative signal
functions in the fits to the $M_{\rm bc}$ distributions and by
varying the end point of the background function~\cite{argus}. 
The uncertainty in modeling the background
shapes in the fits to the $U$ distributions (0.4\% to 3.3\% by mode) 
has contributions from the uncertainties in the simulation of the positron 
and hadron fake rates as well as the input branching fractions in the MC 
simulation. The uncertainty associated with the requirement that there 
be no additional tracks in tagged semileptonic events~(0.3\%) is estimated 
by comparing fully reconstructed $D \bar{D}$ events in data and MC. 
The uncertainty in the semileptonic reconstruction efficiencies due to
imperfect knowledge of the semileptonic form factors 
is estimated by varying the form factors in the MC simulation within
their uncertainties (1.0\%) for all modes except $D^+ \rightarrow
\rho^0 e^+ \nu_e$ and $D^+ \rightarrow \omega e^+ \nu_e$;  for
these a conservative uncertainty~(3.0\%) is taken, as no
experimental information on the form factors in Cabibbo-suppressed
pseudoscalar-to-vector transitions exists.
The uncertainty associated with the simulation of FSR and 
bremsstrahlung radiation in the detector material~(0.6\%) 
is estimated by varying 
the amount of FSR modeled by the PHOTOS algorithm~\cite{PHOTOS}
and by repeating the analysis with and without recovery of photons
radiated by the positron.
The uncertainty associated with the simulation of initial
state radiation~($e^+ e^- \rightarrow D \bar{D} \gamma$) is 
negligible. There is a systematic uncertainty due to finite 
MC statistics~(0.7\% to 4.0\% by mode).

Non-resonant semileptonic decays  $D^+ \to K^- \pi^+ e^+ \nu_e$ 
are background to $\dpkste$. 
There is evidence from the FOCUS experiment for a non-resonant 
component consistent with an S-wave amplitude interfering with 
$\dpkste$~\cite{focusSWave}.
Its contribution, estimated to be 2.4\% in this analysis,
is subtracted in the calculation of the branching fraction of 
$\dpkste$~\cite{nonResSearch}. 
Systematic uncertainties associated with the subtraction are due to 
imperfect knowledge of the amplitude and phase of the non-resonant 
component~(1.0\%), and its effect on the reconstruction 
efficiency~(1.5\%)~\cite{swave_effs}. The lineshapes for semileptonic 
modes with wide resonances are simulated using 
a relativistic Breit-Wigner with a Blatt-Weisskopf form factor. 
A systematic uncertainty associated with the 
$\bar{K}^{*0}$ lineshape~(1.2\%) is assigned by comparing the ($K^- \pi^+$) 
invariant mass distribution in the data to alternative lineshapes and 
the non-resonant contribution. For  $\dprhoe$, there is 
insufficient data to constrain the non-resonant background 
or the resonance lineshape.
The systematic uncertainties from these two sources are expected
to be much smaller than the current statistical uncertainty for this 
mode, and are neglected.

These estimates of systematic uncertainty are added in quadrature
to obtain the total systematic uncertainty~(Table~\ref{table2}): 
4.2\%, 5.6\%, 4.1\%, 6.2\%, and 7.8\% for $\dpke$, $\dppie$, 
$\dpkste$, $\dprhoe$, and $\dpome$, respectively.

We now discuss the results presented in this Letter and 
the $D^0$ semileptonic study in Ref.~\cite{cleoc-neutral_semilep}.
The measured equality of the
inclusive semileptonic widths of $D^0$ and $D^+$ mesons
demonstrates that the source of the lifetime difference
between them is attributable to differences in the hadronic
widths. The widths of the isospin conjugate
exclusive semileptonic decay modes of the $D^0$ and $D^+$
are related by isospin invariance of the hadronic
current. The results obtained here and in Ref.~\cite{cleoc-neutral_semilep}
allow the most precise tests so far.

The ratio $\frac{\Gamma(D^0 \rightarrow K^- e^+ \nu_e)}
{\Gamma(D^+ \rightarrow \bar{K}^0 e^+ \nu_e)}$  is expected
to be unity.
The world average value
is $1.35 \pm 0.19$~\cite{PDG}.
Using our results and the lifetimes of the $D^0$ and
$D^+$~\cite{PDG}, we obtain: 
$
\frac{\Gamma(D^0 \rightarrow K^-e^+\nu_e)} {\Gamma(D^+ \rightarrow
\bar{K}^0 e^+ \nu_e)} = 1.00 \pm 0.05 {\rm (stat)} \pm 0.04 {\rm (syst)}.
$
The result is consistent with unity and with two recent
less precise results: a measurement from BES II using the same
technique~\cite{BESII_ratio} and an indirect measurement from
FOCUS~\cite{FOCUS_ratio,FOCUS_ksmunu}. Ratios of isospin
conjugate decay widths for other semileptonic decay modes are
given in Table~\ref{ratios}.

\begin{table}[ht]
\begin{center}
\begin{tabular}{l c }
\hline
\hline
 \hspace{2.0cm}  Ratio \hspace{2.5cm} & Measured Value \\
\hline
${\Gamma(D^0 \rightarrow K^- e^+ \nu)}/{\Gamma(D^+ \rightarrow \bar{K}^0 e^+ \nu)}$
    & $1.00 \pm 0.05 \pm 0.04 $ \\
${\Gamma(D^0 \rightarrow \pi^- e^+ \nu)}/{[ 2 \cdot \Gamma(D^+ \rightarrow \pi^0 e^+ \nu) ] }$
    & $0.75^{+0.14}_{-0.11} \pm 0.04$ \\
${\Gamma(D^0 \rightarrow K^{*-} e^+ \nu)}/{\Gamma(D^+ \rightarrow  \bar{K}^{*0} e^+ \nu)}$
    & $0.98 \pm 0.08 \pm 0.04$ \\
${\Gamma(D^0 \rightarrow  \rho^{-} e^+ \nu)}/{[ 2 \cdot \Gamma(D^+ \rightarrow \rho^{0} e^+ \nu) ]}$
    & $1.2^{+0.4}_{-0.3} \pm 0.1$ \\
\hline
\hline
\end{tabular}
\end{center}
\caption{ Ratios of semileptonic decay widths of $D^0$
          and $D^+$ mesons. The uncertainties are statistical and systematic.
          In each case the ratio is expected to be unity.  }
\label{ratios}
\end{table}

As the data are consistent with isospin invariance, the precision
of each branching fraction can be improved by
averaging the $D^0$ and $D^+$ results for isospin conjugate
pairs. The isospin-averaged semileptonic 
decay widths, with correlations among systematic uncertainties
taken into account, are given in Table~\ref{decayWidths}.

\begin{table}[ht]
\begin{center}
\begin{tabular}{ l  c }
\hline \hline
 Decay Mode \hspace{10mm} & \hspace{5mm} $\Gamma$~($10^{-2}\cdot{\rm ps}^{-1}$) \hspace{5mm} \\
\hline
$D \rightarrow  K \; e^+ \nu_e  $     & $8.38  \pm   0.20 \pm  0.23 $  \\
$D^0 \rightarrow  \pi^- e^+ \nu_e $   & $0.68 \pm  0.05 \pm 0.02 $  \\
$D \rightarrow  K^* \; e^+ \nu_e $    & $5.32  \pm   0.21 \pm  0.20 $  \\
$D^0 \rightarrow  \rho^- e^+ \nu_e$   & $0.43 \pm   0.06 \pm 0.02$  \\
\hline \hline
\end{tabular}
\end{center}
\caption{Isospin-averaged semileptonic 
         decay widths with
         statistical and systematic uncertainties.
         For Cabibbo-suppressed modes,
         the isospin average is calculated for the $D^0$
         using $\Gamma(D^0) = 2 \cdot \Gamma(D^+)$. }
\label{decayWidths}
\end{table}

The ratio of decay widths for $D \rightarrow \pi  e^+ \nu$ and $D
\rightarrow K e^+ \nu$ provides a test of the LQCD charm 
semileptonic rate ratio prediction~\cite{unquenched_LQCD}. 
Using the isospin-averaged results in
Table~\ref{decayWidths}, we find $\frac{\Gamma(D^0 \rightarrow
\pi^- e^+ \nu)} {\Gamma(D \rightarrow K e^+ \nu)} = (8.1 \pm 0.7
{\rm (stat)}  \pm 0.2 {\rm (syst)} ) \times 10^{-2}$, consistent
with LQCD and two recent results~\cite{pikenu_cleo,pikenu_focus}. 
Furthermore, the ratio $\frac{\Gamma(D
\rightarrow K^* e^+ \nu)} {\Gamma(D \rightarrow K e^+ \nu)}$ is
predicted to be in the range 0.5 to 1.1 (for a compilation see
Ref.~\cite{FOCUS_ratio}). Using the isospin averages in Table~\ref{decayWidths}, 
we find $\frac{\Gamma(D \rightarrow K^* e^+ \nu)} {\Gamma(D
\rightarrow K e^+ \nu)} = 0.63 \pm 0.03 {\rm (stat)} \pm 0.02 {\rm
(syst)}$.

Finally, summing all CLEO-c exclusive semileptonic branching fractions
gives $\sum \mathcal{B}(D^0_{\rm excl}) = (6.1 \pm 0.2 {\rm (stat)}
\pm 0.2 {\rm (syst)} )$\% and $\sum \mathcal{B}(D^+_{\rm excl}) = 
(15.1 \pm 0.5 {\rm (stat)} \pm 0.5 {\rm (syst)})$\%.
These are consistent with 
the world average inclusive semileptonic branching fractions:
$\mathcal{B}(D^0 \rightarrow e^+ X   ) = (6.9 \pm 0.3)$\% 
and $\mathcal{B}(D^+ \rightarrow e^+ X  )
= (17.2 \pm 1.9)$\%~\cite{PDG}, 
excluding the possibility  of additional semileptonic modes 
of the $D^0$ and $D^+$ with large branching fractions.

In summary, we have presented the most precise measurements to
date of the absolute branching fractions
of $D^+$ decays to  $\bar{K}^0 e^+ \nu_e$, $\pi^0 e^+ \nu_e$, 
$\bar{K}^{*0} e^+ \nu_e$, and $\rho^0 e^+ \nu_e$, and the first 
observation and absolute branching fraction measurement of $\dpome$.
We have combined these with 
measurements 
in Ref.~\cite{cleoc-neutral_semilep},
which use the same data and analysis technique,
to demonstrate that charm exclusive semileptonic decays are
consistent with isospin invariance and to test other 
theoretical predictions. 
A comparison of the world average inclusive
semileptonic branching fractions to the sum of the
semileptonic branching fractions in this work 
excludes the possibility of additional semileptonic modes 
with large branching fractions.

The precision achieved in this analysis is consistent with the
expected performance of CLEO-c. CESR is currently running to
collect a much larger $\psi(3770)$ data sample.  It is expected
that this sample will result in greatly improved measurements 
of $D^{0}$ and $D^{+}$ semileptonic branching fractions, 
measurements of the decay form factors, which are stringent 
tests of LQCD, and the CKM matrix elements $\vcs$ and
$\vcd$~\cite{cleoc}. 

We gratefully acknowledge the effort of the CESR staff in
providing us with excellent luminosity and running conditions.
This work was supported by the National Science Foundation and the
U.S. Department of Energy.


\begin{thebibliography}{99}
\bibitem{rhoenu} 

$D^+ \rightarrow  \rho^0 e^+ \nu$ has a small peaking background 
contribution from $D^+ \rightarrow \omega(\pi^- \pi^+)e^+ \nu$ 
that is accounted for in its branching fraction measurement. 
Non-resonant decays contribute background to semileptonic 
decays with vector mesons in the final states and are accounted 
for as discussed below.
\bibitem{nonResSearch}
   The yield of $D^+ \to (K^- \pi^+)_{\rm non-res.} e^+ \nu_e$  measured 
   in bins of the $(K^- \pi^+)$ invariant mass in the data is consistent 
   with this estimate.
\bibitem{FOCUS_ksmunu} FOCUS Collaboration measures
$\Gamma(D^+ \rightarrow \bar{K}^{*0} \mu^+\nu_{\mu})/$
$\Gamma(D^+ \rightarrow \bar{K}^0 \mu^+ \nu_{\mu})$
and $\Gamma(D^+ \rightarrow \bar{K}^{*0} \mu^+\nu_{\mu})/$
$\Gamma(D^+ \rightarrow K^- \pi^+ \pi^+)$. Using
the world average value of
${\cal B} (D^+ \rightarrow K^- \pi^+ \pi^+)$ they extract
$ {\cal B} (D^+ \rightarrow \bar{K}^0 \mu^+ \nu_{\mu})$
and find a value much larger than that in the PDG.
\end{thebibliography}
\end{document}